\documentclass[aps,twocolumn,preprintnumbers,nofootinbib,superscriptaddress]{revtex4-1}
\usepackage{amsmath} \usepackage{graphicx} \usepackage{amsfonts}
\usepackage{array} \usepackage{amsthm} \usepackage{bm}
\usepackage{palatino} \usepackage{mathpazo} 
\usepackage{supertabular}

\usepackage[breaklinks]{hyperref}
\usepackage{color}

\newcommand{\be}{\begin{equation}}
\newcommand{\ee}{\end{equation}}
\newcommand{\ba}{\begin{eqnarray}}
\newcommand{\ea}{\end{eqnarray}}
\newcommand{\bal}{\begin{align}}
\newcommand{\eal}{\end{align}}

\newcommand{\bw}{\begin{widetext}}
\newcommand{\ew}{\end{widetext}}

\begin{document}
\title{Cylindrically symmetric static $n$-dimensional (un)charged (anti-)de Sitter black holes in generic $f(T)$ gravity}

\author{Mustapha Azreg-A\"{\i}nou}
\affiliation{Ba\c{s}kent University, Engineering Faculty, Ba\u{g}l\i ca Campus, Ankara, Turkey}


\begin{abstract}
\noindent \textbf{Abstract:} Given a generic function $f(T)$ we construct in closed forms cylindrically symmetric static $n$-dimensional uncharged and charged de Sitter and anti-de Sitter solutions (including black holes, wormholes and possibly other regular solutions) in $f(T)$ gravity. Applications to some known models are considered.

\vspace{1mm}
\end{abstract}


\maketitle

\section{Introduction}
Black hole solutions endowed with cylindrical symmetry to quadratic and cubic models of $f(T)$ gravity under very restricted conditions are known in the literature~\cite{NS,Awad,Y1,Y2}. Here $T$ is the torsion of the $n$-dimensional spacetime the mathematical definition of which, along with the associated tensors, will be given shortly later in this section. The purpose of this work is to tackle the generic case where both the function $f(T)$ and the dimension $n$ of spacetime are unspecified.

We first consider the uncharged case and provide full details of the analysis yielding the most general solution in closed form for all $f(T)$. Then we provide two examples of applications for quadratic and cubic $f(T)$ with no constraints on the parameters of $f(T)$. We will show that the restrictions introduced on the parameters of $f(T)$, which have been made earlier by some works, have masked some features of the solutions.

Second, we solve in full detail the charged case in almost closed form for any $f(T)$. The final closed form of the solutions depends on the solution of some algebraic or transcendental equations. Again an application to the quadratic case is considered with no restriction on the parameters of $f(T)$ nor on the cosmological constant.

Consider the Lagrangian~\cite{Bengochea,FF7,Sym}
\begin{equation}\label{1}
	{\cal L}=\frac{1}{2\kappa}\int e\big(f(T)+2\Lambda\big)~{\rm d}^{n}x+\int e{\cal L}_{\text{mat}}~{\rm d}^{n}x,
\end{equation}
where ${\cal L}_{\text{mat}}$ is the matter term, $\Lambda$ is the $n$-dimensional cosmological constant
\begin{equation}\label{2}
	\Lambda=\frac{(n-1)(n-2)}{2\ell^2}~\epsilon,\qquad\epsilon=\pm 1,\qquad n\geq 4,
\end{equation}
and $f(T)$ includes nonlinear terms in the torsion $T$. For a self-contained review on $f(T)$ gravity see Ref.~\cite{Maluf}.

This Lagrangian generalizes that of the teleparallel equivalent of general relativity (TEGR)~\cite{HS9}, which uses the vielbein vector fields $e_a=e_{a}{^{\mu}}\partial_\mu$ as fundamental variables instead of the metric $g_{\mu\nu}$, related to each other by
\begin{equation}\label{4}
	g_{\mu \nu} =
	\eta_{ab}e^{a}{_{\mu}}e^{b}{_{\nu}},\quad g^{\mu \nu} =
	\eta^{ab}e_{a}{^{\mu}}e_{b}{^{\nu}},\quad e=\sqrt{|g|},
\end{equation}
with $\eta_{ab}=(+,-,-,-,
\cdots)$ being the metric of the $n$-dimensional Minkowski spacetime and $e\equiv |\det(e^{a}{_{\mu}})|$. 

In cylindrical coordinates ($t$, $r$, $\phi_3$, $\phi_4$ $\cdots$ $\phi_k$, $z_{k+1}$, $z_{k+2}$ $\cdots$ $z_n$), to describe static solutions following Refs.~\cite{NS,Y1,Y1b} we introduce the diagonal vielbein
\begin{equation}\label{5}
	\big(e^{a}{_{\mu}}\big)=\text{diag}\Big(\sqrt{A(r)},\,\frac{1}{\sqrt{B(r)}},\,\underbrace{r,\,r\cdots r}_{k-2\text{ terms}},\,\underbrace{\frac{r}{\ell},\,\frac{r}{\ell}\cdots \frac{r}{\ell}}_{n-k\text{ terms}}\Big),
\end{equation}
resulting in the metric
\begin{equation}\label{m}
	{\rm d}s^2=A(r){\rm d}t^2-\frac{1}{B(r)}{\rm d}r^2-r^2\Big(\sum_{i=3}^{k}{\rm d}\phi^2_i+\sum_{i=k+1}^{n}{{\rm d}z^2_i\over \ell^2}\Big).
\end{equation}
Here $k-2$ is the number of angular coordinates $\phi_i$ ($i: 3\to k$). Once the metric is known one can evaluate the torsion $T$ defined\footnote{$S^{\alpha\mu\nu}$ may be given in a more compact form as:$$S^{\alpha\mu\nu}=\frac{1}{4}(T^{\nu\mu\alpha}+T^{\alpha\mu\alpha}-T^{\mu\nu\alpha})-\frac{1}{2}g^{\alpha\nu}{T^{\sigma\mu}}_{\sigma}+\frac{1}{2}g^{\alpha\mu}{T^{\sigma\nu}}_{\sigma}.$$} by~\cite{Awad,JG}
\begin{align}\label{T0}
	&{T^\alpha}_{\mu \nu}={\Gamma^\alpha}_{\nu \mu}-{\Gamma^\alpha}_{\mu \nu}={e_b}^\alpha
	(\partial_\mu{e^b}_\nu-\partial_\nu{e^b}_\mu),\nonumber\\
	&K_{\alpha\mu\nu}=\frac{1}{2}~(T_{\mu\alpha\nu}+T_{\nu\alpha\mu}-T_{\alpha\mu\nu}),\nonumber\\
	&S^{\alpha\mu\nu}=\frac{1}{2}~(K^{\mu\nu\alpha}-g^{\alpha\nu}{T^{\sigma\mu}}_{\sigma}+g^{\alpha\mu}{T^{\sigma\nu}}_{\sigma}),\nonumber\\
	&T=T_{\alpha\mu\nu}S^{\alpha\mu\nu},
\end{align}
where ${\Gamma^\alpha}_{\mu \nu}=\partial_\nu{e^b}_\mu$ is the Weitzenb\"{o}ck connection~\cite{W}. We\footnote{The expression of $T$ given in Ref.~\cite{Awad} has an extra factor 2 in the term including $A'$.} obtain~\cite{NS}
\begin{equation}\label{T}
	T=\frac{(n-2)A'B}{rA}+\frac{(n-2)(n-3)B}{r^2}.
\end{equation}

It is important to emphasize that in the absence of gravity, the metric~\eqref{m} does not reduce to a Mikowskian metric in cylindrical coordinate; rather, it reduces to the anti-de Sitter solution ($\Lambda>0$) with $A=B=r^2/\ell^2$ in the case $n=4$ and $k=3$~\cite{2b}. In this case, $T$ does not reduce to 0 but to the value of the torsion, $T=6/\ell^2=2\Lambda$, corresponding to a de Sitter or anti-de Sitter spacetime~\cite{book}.

\section{Vacuum solutions}
In the absence of matter, ${\cal L}_{\text{mat}}=0$, the field equations take the following form
\begin{align}
\label{e1}& 2Tf_T-2\Lambda-f=0,\\
\label{e2}& \frac{2(n-2)Bf_{TT} T'}{r}+\frac{(n-2)f_T[2(n-3)AB+rBA'+rAB']}{r^2 A}\nonumber \\
& -f-2\Lambda=0,\\
\label{e3}& \frac{f_{TT} [r^2T+(n-2)(n-3)B]T'}{(n-2)r}-f-2\Lambda
\nonumber \\
& +\frac{f_T}{2r^2{A}^2}\Big[2r^2ABA''-r^2BA'^2+4(n-3)^2A^2B\\
& +2(2n-5)rABA'+r^2AA'B' +2(n-3)rA^2B'\Big]=0,\nonumber
\end{align}
where $T'\equiv\partial T/\partial r$, $f_T\equiv\partial f/\partial T$ and $f_{TT}\equiv\partial^2 f/\partial T^2$. For a \emph{given} function $f(T)$, Eq.~\eqref{e1} is not a differential equation for $f(T)$; rather it is an algebraic equation or a transcendental one for $T$. Thus, in the absence of matter, it follows that the torsion $T$ is constant. This constant depends on $\Lambda$ and on the parameters that enter into the definition of $f(T)$. For short we write $T\equiv T(\Lambda,\cdots)$. Since, necessarily, $f(0)=0$, we conclude from~\eqref{e1} that the value of this constant must be different from 0 if $\Lambda\neq 0$.

Since $T=\text{const}\neq 0$, all terms including $T'$ vanish. Using~\eqref{e1} in~\eqref{e2} and~\eqref{e3} we bring them respectively to
\begin{align}
\label{E1}& (n-2) [2 (n-3)+r (F+G)]=\frac{2 T r^2}{B},\\
\label{E2}& 2 F'+F^2+2 (2 n-5) \frac{F}{r}+F G+2 \frac{(n-3)}{r^2} [2 (n-3)+r G]\nonumber \\
& =\frac{4 T}{B},
\end{align}
where
\begin{equation}
F\equiv {A'\over A},\qquad G\equiv {B'\over B}.
\end{equation}
On eliminating $T/B$ from~\eqref{E1} and~\eqref{E2} we arrive at the master equation
\begin{equation}\label{mas0}
2 r^2 F'+(r G+2 n-6) r F+r^2 F^2-2 r G+12-4 n=0,
\end{equation}
and on eliminating $T/B$ from~\eqref{E1} and~\eqref{T} we arrive at
\begin{equation}\label{mas1}
G=F,
\end{equation}
bringing thus~\eqref{mas0} to
\begin{equation}\label{mas}
r^2 F'+(n-4) r F+r^2 F^2+6-2 n=0.
\end{equation}
Direct integrations yield
\begin{align}
&F=\frac{(3-n) r+2 C_1(n-1) r^n}{r [r+C_1(n-1) r^n]},\\
\label{A}&A=B=C_1 C_2 (n-1) r^2+\frac{C_2}{r^{n-3}},
\end{align}
where ($C_1,\,C_2$) are constants of integration. By~\eqref{mas1} we have $A=\text{const} \,B$ and we can set $\text{const}=1$ on rescaling the time coordinate $t$~\eqref{m}. Following~\cite{Awad} we set $C_2=-m$, where $m$ is the mass parameter, and we fix the value of $C_1$ upon substituting~\eqref{A} into~\eqref{E1}
\begin{equation}\label{C1}
C_1=-\frac{T}{(n-2)(n-1)^2m}.
\end{equation}
Finally the solution reads
\begin{equation}\label{S}
A=B=\frac{T(\Lambda,\cdots)}{(n-2)(n-1)}~r^2-\frac{m}{r^{n-3}},
\end{equation}
where $T(\Lambda,\cdots)$ is the constant value of the torsion.
Hence if the solution of the algebraic or transcendental Eq.~\eqref{e1} provides a positive value for $T(\Lambda,\cdots)$, the metric~\eqref{S} describes an anti-de Sitter solution with an effective cosmological constant
\begin{equation}\label{Lp}
	\Lambda_{\text{eff}}=\frac{3T(\Lambda,\cdots)}{(n-2)(n-1)}>0,
\end{equation}
and if the solution of Eq.~\eqref{e1} provides a negative value for $T(\Lambda,\cdots)$, the metric~\eqref{S} describes a de Sitter solution with an effective cosmological constant
\begin{equation}\label{Ln}
	\Lambda_{\text{eff}}=\frac{3T(\Lambda,\cdots)}{(n-2)(n-1)}<0.
\end{equation}
In these definitions of an anti-de Sitter or a de Sitter solution we mean that the behavior of the solution as $r\to\infty$ is that of an anti-de Sitter or a de Sitter black hole.

As we shall see in the applications, $\Lambda_{\text{eff}}$ and $\Lambda$~\eqref{2} may have different signs. That is, for a given $\epsilon$, $\Lambda_{\text{eff}}$ may have both signs depending on the other parameters entering the definition of $f(T)$.

In the linear case, $f(T)=T$, Eq.~\eqref{e1} yields $T=2\Lambda$ and
\begin{equation}\label{Ll}
\Lambda_{\text{eff}}=\frac{6\Lambda}{(n-2)(n-1)},\qquad\text{for all }\Lambda .
\end{equation}
In 4-dimensional spacetime, $n=4$, we obtain $\Lambda_{\text{eff}}=\Lambda$.

In the case $\Lambda=0$, the algebraic or transcendental Eq.~\eqref{e1} may admit further solutions other than the trivial solution $T=0$. This means that a nonlinear $f(T)$ would always generate an effective cosmological constant.

The solution~\eqref{S} was also derived in~\cite{Awad} but in the \emph{very special case} $f(T)=T+\alpha T^2$ with $24\alpha\Lambda=-1$ and $\alpha <0$. Here we have obtained the solution~\eqref{S} for a generic function $f(T)$ and with no constraint on $\Lambda$.

\subsection*{Applications}
We consider the power-law cases $f(T)=T+\alpha T^2$ and $f(T)=T+\alpha T^2+\beta T^3$, respectively, which offer the models in good agreement with observational data~\cite{d0,d1,d2,d3}.

\subsubsection{$f(T)=T+\alpha T^2$}
For $f(T)$ including a quadratic torsion term
\begin{equation}\label{3}
	f(T)=T+\alpha T^2,\qquad\alpha\in\mathbb{R},
\end{equation}
we obtain solving~\eqref{e1}, which takes the form
\begin{equation}
	3\alpha T^2+T-2\Lambda =0,
\end{equation}
two real solutions
\begin{equation}\label{T12}
	T_{\pm}=-\frac{1\pm \sqrt{1+24\alpha\Lambda}}{6\alpha},
\end{equation}
provided
\begin{equation}\label{prvd}
	24\alpha\Lambda\geq -1.
\end{equation}
This constraint is always satisfied if $\alpha$ and $\Lambda$ have the same sign. Thus,
{\small
	\begin{align}\label{con}
		&\alpha>0,\;\Lambda>0: \left\{
		\begin{array}{ll}
			T_+<0, & \hbox{$T=T_+\Rightarrow$ \text{de Sitter}} \\
			T_->0, & \hbox{$T=T_-\Rightarrow$ \text{anti-de Sitter}}
		\end{array}
		\right.\nonumber\\
		&\alpha<0,\;\Lambda<0: \left\{
		\begin{array}{ll}
			T_+>0, & \hbox{$T=T_+\Rightarrow$ \text{anti-de Sitter}} \\
			T_-<0, & \hbox{$T=T_-\Rightarrow$ \text{de Sitter}}
		\end{array}
		\right.\nonumber\\
		&\alpha>0,\;\Lambda<0,\;24\alpha\Lambda\geq -1: \left\{
		\begin{array}{ll}
			T_+<0, & \hbox{$T=T_+\Rightarrow$ \text{de Sitter}} \\
			T_-<0, & \hbox{$T=T_-\Rightarrow$ \text{de Sitter}}
		\end{array}
		\right.\nonumber\\
		&\alpha<0,\;\Lambda>0,\;24\alpha\Lambda\geq -1: \left\{
		\begin{array}{ll}
			T_+>0, & \hbox{$T=T_+\Rightarrow$ \text{anti-de Sitter}} \\
			T_->0, & \hbox{$T=T_-\Rightarrow$ \text{anti-de Sitter}}
		\end{array}
		\right.\nonumber\\
		&\Lambda=0: \left\{
		\begin{array}{ll}
			\alpha>0\Rightarrow T_+<0, & \hbox{$T=T_+\Rightarrow$ \text{de Sitter}} \\
			\alpha<0\Rightarrow T_+>0, & \hbox{$T=T_+\Rightarrow$ \text{anti-de Sitter}.}
		\end{array}
		\right.
\end{align}}
The solution is given in~\eqref{S} on replacing $T(\Lambda,\cdots)$ by $T_{\pm}$~\eqref{T12}. We see that when $\alpha$ and $\Lambda$ have opposite signs, it is the sign of $\Lambda$ that determines the nature of the solution: For $\Lambda<0$ the solution is of the de Sitter type and for $\Lambda>0$ the solution is of the anti-de Sitter type.

In the limit $\alpha\to 0$, we have $\lim_{\alpha\to 0}T_-=2\Lambda$ and we recover the result~\eqref{Ll} of general relativity.

In Ref.~\cite{Awad} only the special case $24\alpha\Lambda = -1$, $\alpha<0$, and $f(T)=T+\alpha T^2$ has been considered.

\subsubsection{$f(T)=T+\alpha T^2+\beta T^3$}
This case is more involved and we will not consider it in full detail. Equation~\eqref{e1} reduces to
\begin{equation}\label{gen}
	5\beta T^3+3\alpha T^2+T-2\Lambda =0,
\end{equation}
then to the Weierstrass polynomial on eliminating the quadratic term
\begin{align}\label{W}
	&4z^3-g_2z-g_3=0,\qquad T=z-\frac{\alpha}{5\beta},\\
	&g_2=\frac{4 (3 \alpha ^2-5 \beta)}{25 \beta ^2},\qquad
	g_3=\frac{4 (50 \beta ^2 \Lambda +5 \alpha  \beta -2 \alpha ^3)}{125 \beta ^3}.\nonumber
\end{align}
A complete description on how to determine the roots of the Weierstrass polynomial is given in~\cite{W1} and in Appendix A of~\cite{W2}. The number of the real roots depend on the signs of $g_2$ and $\Delta\equiv g_2^3-27g_3^2$. For $g_2>0$ and $\Delta>0$ there are three distinct real roots; for $g_2>0$ and $\Delta =0$ there are two distinct real roots; and for $\Delta <0$ there is one real root.

As we noticed earlier even in the case $\Lambda\equiv 0$,  the $f(T)$ theory generates a non vanishing effective cosmological constant. Let us examine this case which is much easier than the generic case~\eqref{gen}. Solutions to
\begin{equation}\label{gen0}
	5\beta T^3+3\alpha T^2+T =0,
\end{equation}
include the trivial one $T_0=0$ and
\begin{equation}\label{Tb12}
	T_{\pm}=-\frac{3\alpha}{10\beta}\bigg(1\pm \sqrt{1-\frac{20\beta}{9\alpha^2}}~\bigg)=-\frac{3\alpha\pm\sqrt{9\alpha^2-20\beta}}{10\beta}.
\end{equation}
This expression of $T_{\pm}$ was obtained in~\cite{NS} as a by-product without fully treating the case $\Lambda=0$. This yields the following cases.
{\small
	\begin{align}\label{con1}
		&\alpha<0,\;\beta<0: \left\{
		\begin{array}{ll}
			T_+<0, & \hbox{$T=T_+\Rightarrow$ \text{de Sitter}} \\
			T_->0, & \hbox{$T=T_-\Rightarrow$ \text{anti-de Sitter}}
		\end{array}
		\right.\nonumber\\
		&\alpha\geq 0,\;\beta<0: \left\{
		\begin{array}{ll}
			T_+>0, & \hbox{$T=T_+\Rightarrow$ \text{anti-de Sitter}} \\
			T_-<0, & \hbox{$T=T_-\Rightarrow$ \text{de Sitter}}
		\end{array}
		\right.\nonumber\\
		&\alpha>0,\;9\alpha^2\geq 20\beta>0: \left\{
		\begin{array}{ll}
			T_+<0, & \hbox{$T=T_+\Rightarrow$ \text{de Sitter}} \\
			T_-<0, & \hbox{$T=T_-\Rightarrow$ \text{de Sitter}}
		\end{array}
		\right.\nonumber\\
		&\alpha<0,\;9\alpha^2\geq 20\beta>0: \left\{
		\begin{array}{ll}
			T_+>0, & \hbox{$T=T_+\Rightarrow$ \text{anti-de Sitter}} \\
			T_->0, & \hbox{$T=T_-\Rightarrow$ \text{anti-de Sitter}.}
		\end{array}
		\right. .
\end{align}}
The solution is given in~\eqref{S} on replacing $T(\Lambda,\cdots)$ by $T_{\pm}$~\eqref{Tb12}. We see that when $\beta$ is positive we have either a de Sitter solution or an anti-de Sitter solution, but not both. When $\beta$ is negative we may have both de Sitter and anti-de Sitter solutions.

\section{Charged solutions}
Now, let us go back to the generic case where $f(T)$ is not necessarily quadratic or cubic.

In the presence of an electromagnetic source ${\cal L}_{\text{mat}}={\cal L}_{\text{em}}$, corresponding to an energy-momentum tensor $$T_{\text{(em)}\,\mu}{}^\nu=F_{\mu \alpha}F^{\nu \alpha}-\frac{1}{4} \delta_\mu{}^\nu F_{\alpha \beta}F^{\alpha \beta},$$ with $F_{\alpha \beta}=\partial_\alpha A_\beta-\partial_\beta A_\alpha$, $T$ is no longer a constant; rather, $T\equiv T(r)$. Solutions endowed with cylindrical symmetry have their vector potential given by $A_t=V(r)$ with the remaining components being 0. Taking $\kappa=-2$, the field equations read
\begin{align}
	\label{c1}& 2Tf_T-2\Lambda-f+\frac{2B}{A}~V'^2=0,\\
	\label{c2}& \frac{2(n-2)Bf_{TT} T'}{r}+\frac{(n-2)f_T[2(n-3)AB+rBA'+rAB']}{r^2 A}\nonumber \\
	& -f-2\Lambda+\frac{2B}{A}~V'^2=0,\\
	\label{c3}& \frac{f_{TT} [r^2T+(n-2)(n-3)B]T'}{(n-2)r}-f-2\Lambda -\frac{2B}{A}~V'^2
	\nonumber \\
	& +\frac{f_T}{2r^2{A}^2}\Big[2r^2ABA''-r^2BA'^2+4(n-3)^2A^2B\\
	& +2(2n-5)rABA'+r^2AA'B' +2(n-3)rA^2B'\Big]=0,\nonumber\\
	\label{c4}&\partial_r(\sqrt{|g|}F^{\mu r})=0,
\end{align}
where $V'\equiv\partial V/\partial r$. On eliminating $\Lambda$ between~\eqref{c1} and~\eqref{c2} we arrive at
\begin{equation}\label{c5}
	2 (n-2)(\ln  f_T)' r+(n-2) [2 (n-3)+r (F+G)]=\frac{2 T r^2}{B},
\end{equation}
where $(\ln  f_T)'=\partial _r (\ln  f_T)$. On eliminating $T/B$ between~\eqref{c5} and~\eqref{T} we arrive at
\begin{equation}\label{c5b}
	G=F-2(\ln  f_T)',
\end{equation}
which implies
\begin{equation}\label{c6}
	B(r)=\Big[\frac{f_T(T_0)}{f_T(T(r))}\Big]^2~A(r),
\end{equation}
where the constant $T_0\equiv\lim_{r\to\infty}T(r)$. As we shall see shortly later in this section, Eq.~\eqref{c7}, the function $V'$ vanishes in the limit $r\to\infty$, consequently the last term in~\eqref{c1} drops to 0 in this limit and the equation reduces to~\eqref{e1}. Said otherwise, for a given function $f(T)$, $T_0$ is a root of the algebraic or transcendental Eq.~\eqref{e1}. Notice that both functions $A$ and $B$ must have the same number of zeros since $f_T(T_0)/f_T(T(r))\neq 0$ and $f_T(T(r))/f_T(T_0)\neq 0$. For instance, if $A$ has no zero, the solution is either a non-horizon solution, a wormhole, or a regular solution~\cite{Rod}.

The next step is to reduce~\eqref{c4}. Using~\eqref{c6} we obtain
\begin{equation}
	\sqrt{|g|}=\frac{f_T(T(r))}{f_T(T_0)}~r^{n-2}\ell^{k-n},
\end{equation}
where $k-2$ is the number of angular coordinates~\eqref{m}, and
\begin{equation}
	F^{tr}=\frac{B}{A}~V'=\Big[\frac{f_T(T_0)}{f_T(T(r))}\Big]^2~V'.
\end{equation}
Using these last two equations and introducing the charge parameter $q$ we integrate~\eqref{c4} by
\begin{equation}\label{c7}
	V'=-\frac{f_T(T(r))}{f_T(T_0)}~\frac{(n-3)q}{r^{n-2}}.
\end{equation}
Since by definition $T_0\equiv\lim_{r\to\infty}T(r)$, we see that $V'\to -(n-3)q/r^{n-2}\to 0$ as $r\to\infty$.

An algorithm for solving the field equations~\eqref{c1} to~\eqref{c4} for a \emph{given} function $f(T)$ consists in the following five steps.
\begin{enumerate}
	\item Solve~\eqref{e1} for $T_0$. Insert~\eqref{c6} and~\eqref{c7} into~\eqref{c1} to obtain
	\begin{equation}
		2Tf_T(r)-2\Lambda-f+\frac{2(n-3)^2q^2}{r^{2(n-2)}}=0,
	\end{equation}
	Solve this algebraic equation for $T(r)$ and write $T(r)$ as $T(r)=T_0+\mathcal{T}(r)$ with $\lim_{r\to\infty}\mathcal{T}(r)=0$;
	\item  Insert~\eqref{c6} into~\eqref{T} to obtain,
	\begin{equation}
		T(r)=\Big[\frac{f_T(T_0)}{f_T(T(r))}\Big]^2~\Big[\frac{(n-2)}{r}~A'+\frac{(n-2)(n-3)}{r^2}~A\Big],
	\end{equation}
	and solve this first order differential equation for $A(r)$ or $\mathcal{A}(r)$ where
	\begin{equation}
		A(r)=A_0(r)+\mathcal{A}(r),
	\end{equation}
	with $\lim_{r\to\infty}\mathcal{A}(r)=0$ and $A_0(r)$ is the uncharged solution~\eqref{S}. We draw the following interesting conclusion: Since $\lim_{r\to\infty}\mathcal{A}(r)=0$, the only term containing $r^2$ is in $A_0(r)$ and this means that it is the constant torsion $T_0$ that generates an effective cosmological constant for charged solutions as is the case for uncharged solutions. Moreover, this effective cosmological constant has the same value for the charged and uncharged solutions;
	\item Obtain $B(r)$ from~\eqref{c6}
	\begin{equation}
		B(r)=A_0(r)+\Big[\frac{f_T(T_0)^2}{f_T(T(r))^2}-1\Big]A_0(r)+\Big[\frac{f_T(T_0)^2}{f_T(T(r))^2}\Big]\mathcal{A}(r);
	\end{equation}
	\item Obtain $V(r)$ from~\eqref{c7}
	\begin{equation}
		V=\frac{q}{r^{n-3}}-(n-3)q\int^{r}\Big[\frac{f_T(T(r))}{f_T(T_0)}-1\Big]~\frac{{\rm d}r}{r^{n-2}};
	\end{equation}
	\item Use~\eqref{c3} for checking consistency of the results. The latter is brought to the following form once we eliminate $T$, $\Lambda$ and $G$ using~\eqref{T}, \eqref{c1} and\eqref{c5b}
	\begin{equation}\label{cons}
		r^2 F'+(n-4) r F+r^2 F^2+6-2 n=\frac{4r^2V'^2}{f_T(T(r))A}.
	\end{equation}
	Compare with\eqref{mas}.
\end{enumerate}
Since $\lim_{r\to\infty}A(r)=\lim_{r\to\infty}B(r)=A_0(r)$, the classification of the solutions (de Sitter or anti-de Sitter) follows that of the uncharged solution. For instance for $f(T)=T+\alpha T^2$, the classification of the solutions is that given in~\eqref{con} where $T_{\pm}$ is now denoted by $T_0$ for short.

\subsection*{Application}
We provide an example (not treated in the literature) on how to apply the five previously described processing steps. We consider again the case $f(T)=T+\alpha T^2$. Equation~\eqref{c1} reads
\begin{equation}
	3\alpha T^2+T-2\Lambda +\frac{2(n-3)^2q^2}{r^{2(n-2)}}=0.
\end{equation}
For illustration we restrict ourselves to one of the two roots of this equation fixing $n=4$.
\begin{align}\label{a1}
	&T(r)=-\frac{1- \sqrt{1+24\alpha\lambda(r)}}{6\alpha},\quad\lambda(r)\equiv \Lambda-\frac{q^2}{r^{4}},\nonumber\\
	&T(r)=T_0+\frac{\sqrt{1+24\alpha\lambda(r)}-p}{6\alpha},\quad p\equiv \sqrt{1+24\alpha\Lambda},
\end{align}
where $T_0=(p-1)/(6\alpha)$~\eqref{T12}. This is no restriction, for $T_0$ still can have both signs satisfying $T_0\Lambda<0$ or $T_0\Lambda\alpha>0$. Thus, we will have both de Sitter and anti-de Sitter charged solutions. Moreover, the solution corresponding to the other root of $T_0$, $-(1+p)/(6\alpha)$, is derived from that corresponding to the root $T_0=(p-1)/(6\alpha)$ on replacing $p$ by $-p$. However, as we shall see below, the features of these solutions are different.
\begin{figure*}
	\centering
	\includegraphics[width=0.49\textwidth]{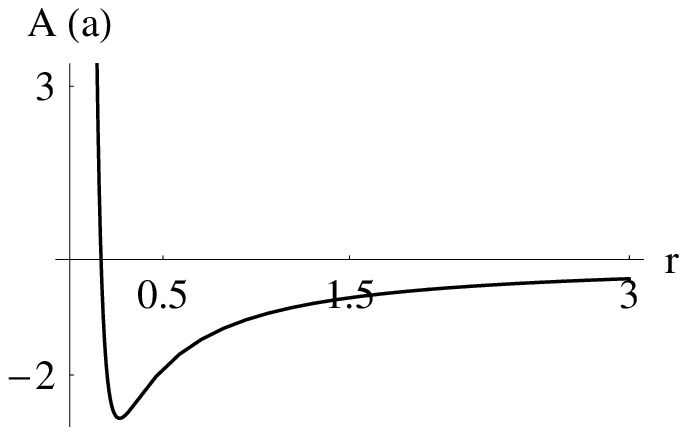}
	\includegraphics[width=0.49\textwidth]{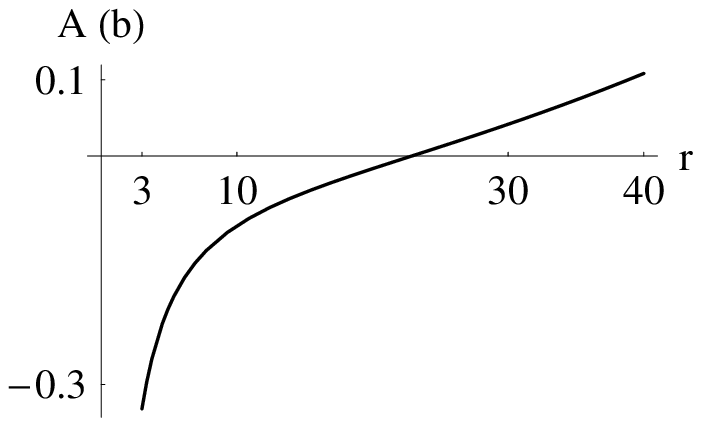} \\
	\caption{\footnotesize{Plot of $A(r)$  for $p = 0.7$, $m = 1$, $q = 0.1$ and $\alpha = -100$ showing an anti-de Sitter black hole solution. In (a) $0<r\leq 3$ and in (b) $3\leq r<\infty$.}}\label{Fig1}
\end{figure*}
\begin{figure*}
	\centering
	\includegraphics[width=0.49\textwidth]{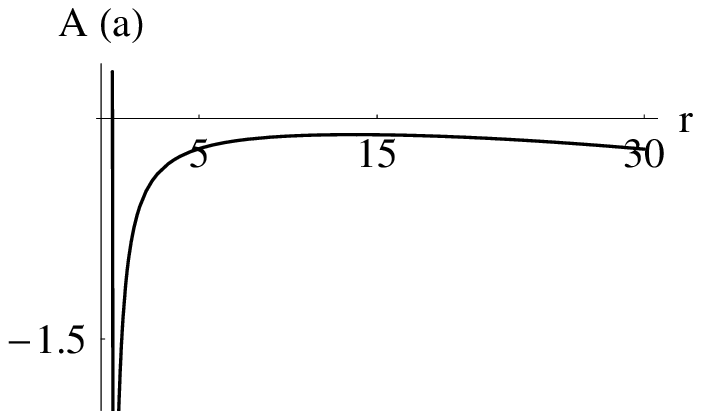}
	\includegraphics[width=0.49\textwidth]{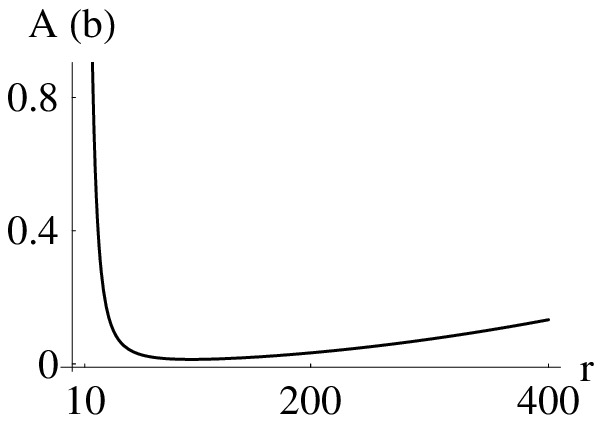} \\
	\caption{\footnotesize{Plot of $A(r)$. (a) $p = 1.7$, $m = 1$, $q = 0.1$ and $\alpha = -100$ showing a de Sitter black hole solution. (b) $p = 0.7$, $m = 0.04$, $q = 10$ and $\alpha = -10000$ showing a naked singularity.}}\label{Fig2}
\end{figure*}
Let
\begin{equation}\label{a2}
	R(r)\equiv \sqrt{p^2r^4-24q^2\alpha},\qquad\alpha<0.
\end{equation}
If $\alpha>0$ the radial coordinate has a minimum value, $r_{\text{min}}=24q^2\alpha/p^2$, and the solution may describe a wormhole if no horizon for $r>r_{\text{min}}$.

Taking $p=0$ ($24\alpha\Lambda=-1$) yields $T_0=-1/(6\alpha)$ (a double root). From~\eqref{a2} we see that the only possibility left is to assume $\alpha<0$ restricting thus the sign of $\Lambda$: $\Lambda>0$.  This very special case has been treated in Ref.~\cite{Awad}, so we will not consider it here. Our purpose in this section is to tackle the most general case for $f(T)=T+\alpha T^2$. In this section we assume $p\neq 0$.

Looking for a solution $A(r)=(T_0/6)r^2-m/r+\mathcal{A}(r)$ we find that $\mathcal{A}$ satisfies
\begin{equation}\label{a3}
	r\mathcal{A}'+\mathcal{A}=\frac{1}{(2+p)^2}\Big[\frac{R[r]^3}{12 \alpha  r^4}-\frac{6 q^2}{r^2}-\frac{p^3r^2}{12 \alpha }\Big],
\end{equation}
where we have used
\begin{equation}\label{a3b}
	\Big[\frac{f_T(T_0)}{f_T(T(r))}\Big]^2=\frac{(2+p)^2r^4}{[2r^2+R(r)]^2}.
\end{equation}
This yields
\begin{align}\label{a4}
	&A=\frac{p-1}{36\alpha}~r^2-\frac{m}{r}+\frac{1}{(2+p)^2}\Big[\frac{6q^2}{r^2}-\frac{p^3r^2}{36\alpha}\nonumber\\
	&+\frac{1}{12\alpha r}\int^r \frac{R(r)^3}{  r^4}~{\rm d}r\Big],\nonumber\\
	&V=\frac{q}{r}-\frac{qp}{(2+p)r}-\frac{q}{2+p}\int^{r}
	\frac{R(r)\,{\rm d}r}{r^4}.
\end{align}
A further integration by parts yields
\begin{align}\label{a4b}
	&A=\frac{p-1}{36\alpha}~r^2-\frac{m}{r}+\frac{1}{(2+p)^2}\Big[\frac{6q^2}{r^2}-\frac{p^3r^2}{36\alpha}-\frac{R(r)^3}{36\alpha r^4}\nonumber\\
	&+\frac{p^2}{6\alpha r}\int^r R(r)~{\rm d}r\Big],\nonumber\\
	&V=\frac{q}{r}-\frac{qp}{(2+p)r}+\frac{qR(r)}{3(2+p)r^3}-\frac{2qp^2}{3(2+p)}\int^{r}
	\frac{{\rm d}r}{R(r)}\nonumber\\
	&B=\frac{(2+p)^2r^4}{[2r^2+R(r)]^2}~A.
\end{align}
The expression of $B$ has been derived from~\eqref{c6} and~\eqref{a3b}. We have checked that this expression of $A$ along with that of $V'$~\eqref{c7} satisfy~\eqref{cons}.

The two integrals in~\eqref{a4b} could be expressed in terms of the incomplete elliptic integral of the first kind~\cite{wolframe}. We content ourselves with the following series formulas
\begin{multline}\label{a5}
	A=\frac{p-1}{36\alpha}~r^2-\frac{m}{r}+\frac{1}{(2+p)^2}\Big[\frac{3 (2+p) q^2}{r^2}-\frac{18 q^4 \alpha }{5 p r^6}\\-\frac{8 q^6 \alpha ^2}{p^3 r^{10}}-\frac{648 q^8 \alpha ^3}{13 p^5 r^{14}}+O\Big(\frac{1}{r^{18}}\Big)\Big],
\end{multline}
\begin{multline}\label{a6}
	V=\frac{q}{r}-\frac{1}{2+p} \Big[\frac{12 q^3 \alpha }{5 p r^5}+\frac{8 q^5 \alpha ^2}{p^3 r^9}\\+\frac{864 q^7 \alpha ^3}{13 p^5
		r^{13}}+O\Big(\frac{1}{r^{17}}\Big)\Big].
\end{multline}

First notice from~\eqref{a5} and~\eqref{a6} that in the limit $\alpha\to 0$ ($p\to 1$ from the left and from the right) we obtain the results of general relativity as the effective cosmological constant $T_0/2=(p-1)/(12\alpha)$ reduces to $\Lambda$. 

Now consider the case $\Lambda=0$. This implies $p=1$, $T_0=0$ and $\Lambda_{\text{eff}}=0$.  In the limit $\Lambda\to 0$ the effective cosmological constant vanishes. Thus, $\Lambda=0$ and for this value of $T_0=(p-1)/(6\alpha)$ the $f(T)$ theory does not generate an effective cosmological constant. Had we chosen the other value of $T_0=-(1+p)/(6\alpha)$ we would have obtained a nonvanishing effective cosmological constant but we would not have recovered the results of general relativity in the limit $\alpha\to 0$. The solution corresponding to $T_0=-(1+p)/(6\alpha)$ is derived from~\eqref{a5} and~\eqref{a6} on replacing $p$ by $-p$. The fact that the results of general relativity are not always recovered in $f(T)$ gravity was noticed in~\cite{Awad}.

We see how the first correction to general relativity, $f(T)=T+\alpha T^2$, already provides solutions not in the realm of general relativity. This applies to charged and uncharged solutions.

The nature of the solution~\eqref{a4b} is revealed upon investigating the sign of the function $A$. The parameter $\alpha$ being negative~\eqref{a2}, the solution is certainly not a wormhole for $r$ cannot have a minimum value. It is obvious from the expression of $A$ that the third, fourth and fifth terms are positive while the second and sixth terms are negative, and the sign of the first term depends on the sign of $p-1$. Since some terms are positive and some other terms are negative, the equation $A=0$ may have \emph{positive} roots. If this is the case, the solution is a multi-horizon anti-de Sitter black hole ($0<p<1$), as can be seen from~\eqref{a5}, or a multi-horizon de Sitter black hole ($p>1$). If $A=0$ admits no \emph{positive} root, then the solution is a naked singularity. Recall that by an anti-de Sitter or a de Sitter solution we mean that the behavior of the solution as $r\to\infty$ is that of an anti-de Sitter or a de Sitter black hole.

Now, as $r\to 0$ the fifth term in~\eqref{a4b} is preponderant and behaves as $+1/r^4\to +\infty$. For $p>1$ the first term in~\eqref{a5} is preponderant and goes to $-\infty$ as $r\to \infty$. Thus, for $p>1$ there is at least one root to $A=0$ and the solution is a black hole. For $p<1$, in order to reveal the nature of the solution, we have to resolve to numerical analysis. Figures~\ref{Fig1} and~\ref{Fig2} depict the graph of $A(r)$ for $p=0.7<1$ showing the existence of an anti-de Sitter black hole and of a naked singularity. A de Sitter black hole solution for $p=1.7>1$ is also shown.\\

\section{Multipole expansion: Generic case}
Let us go back again to the generic case where $f(T)$ is not necessarily quadratic or cubic. The purpose of this section is to determine the first nonvanishing higher order moments of the electric potential $V$, as we did in~\eqref{a6}, but when $f(T)$ is generic fixing $n=4$. Skipping the calculations, we provide the answer setting $x=1/r$:
\begin{widetext}
\begin{align}\label{m1}
&T=T_0-\frac{48 q^2}{f'(T_0)+2 T_0 f''(T_0)}~x^4-\frac{80640 q^4 [3 f''[T_0]+2 T_0 f^{(3)}(T_0)]}{[f'(T_0)+2
	T_0 f''(T_0)]^3}~x^8+ \cdots \nonumber\\
&V=q x-\frac{2 q^3 f''(T_0) }{5 f'(T_0) [f'(T_0)+2 T_0 f''(T_0)]}~x^5 -\frac{2 q^5 [3
	f''(T_0)^2-f'(T_0) f^{(3)}(T_0)]}{9 f'(T_0) [f'(T_0)+2 T_0 f''(T_0)]^3}~x^9+
\cdots ,
\end{align}
\end{widetext}
where $f^{(3)}$ denotes third derivative. \emph{Only in these expressions a prime notation does not denote a derivative with respect to} $r$: All derivatives are evaluated with respect to $T$. For instance: $f^{(3)}\equiv f_{TTT}$ and $f''\equiv f_{TT}$. The term $f'(T_0)$ is simply $[f(T_0)-2 \Lambda]/(2 T_0)$ and, recall, $T_0$ is a solution to the algebraic or transcendental Eq.~\eqref{e1} for a given function $f(T)$. We checked that for $f(T)=T+\alpha T^2$ we reobtain the results of the previous section~\eqref{a6}.

We see that the first nonvanishing term is proportional to $1/r^5$ unless $f''(T_0)=0$ which was never the case in all models considered in the literature.

This generic expansion is not valid in the special case where the denominator in~\eqref{m1} vanishes: $f'(T_0)+2 T_0 f''(T_0)=0$. This is, for instance, the case with $f(T)=T+\alpha T^2$ if $p=0$ ($24\alpha\Lambda=1$), which yields $T_0=-1/(6\alpha)$ (a double root). In this case the first nonvanishing term is proportional to $1/r^3$ if $n=4$~\cite{Awad}.

\section{Conclusion}
Applications of the generic solution to quadratic and cubic cases revealed that the $f(T)$ theory may generate an effective cosmological constant in the absence of a real one $\Lambda=0$. It also generates solutions that do not recover any of the known solutions of general relativity.

To the best of our knowledge, earlier works determined only anti-de Sitter solutions, mainly due to parameters' restrictions. We have shown both end-behaviors, de Sitter and anti-de Sitter, are possible within generic $f(T)$ theory and with known quadratic and cubic models.

When no restrictions are made on the parameters defining the function $f(T)$ and/or the cosmological constant, the $f(T)$ theory reveals further features not encountered in the special cases: a) Generically speaking, the electric potential has its first nonvanishing higher order moment proportional to $1/r^5$. This shows that the shape of $f(T)$ is unimportant as to fix the first nonvanishing higher order term, but important as to fix the value of the moment. b) For both charged and uncharged cases it is the constant torsion $T_0$ that generates an effective cosmological constant which has the same value for the charged and uncharged solutions.


%



\end{document}